\documentclass[aps,preprint,epsfig,rotate]{revtex4}
\usepackage{sansmath,amsmath,pdflscape,bm,epsfig,graphicx,rotating}
\usepackage{pdfpages}
\usepackage{pdflscape}
\newcommand{\be}{\begin{equation}}
\newcommand{\ee}{\end{equation}}
\newcommand{\bea}{\begin{eqnarray}}
\newcommand{\eea}{\end{eqnarray}}
\begin{document}

\title{Hylleraas-Configuration Interaction calculations on the ${\bf 1}^{\bf 1}{\bf S}$ 
ground state of helium atom }

\author{Mar\'{\i}a Bel\'{e}n Ruiz} 
\email[E--mail address: ]{maria.belen.ruiz@fau.de}

\affiliation{Theoretical Chemistry\\
Department of Chemistry and Pharmacy, \\
 Friedrich-Alexander-University Erlangen-N\"urnberg,
Egerlandstra\ss e 3, 91058 Erlangen, Germany}

\date{\today}

\begin{abstract}
Hylleraas-Configuration Interaction (Hy-CI) calculations on the ground $%
1^{1}S $ state of helium atom are presented using $s$-, $p$-, $d$-, and $f$-Slater 
orbitals of both real and complex form. Techniques of construction of
adapted configurations, optimization of the orbital exponents and structure of
the wave function expansion are explored. A new method to evaluate the
two-electron kinetic energy integrals occurring in the Hy-CI method has been  
tested in this work and compared with other methods. 
The non-relativistic Hy-CI energy values are $\approx$ 10 picohartree accurate, about $2.2\times 10^{-6}$  
$\mathrm{cm}^{-1}$. The Hy-CI calculations are compared with Configuration 
Interaction (CI) and Hylleraas (Hy) calculations employing the same orbital basis set, same 
computer code and same computer machines. The computational required times are reported. 
\end{abstract}


\vspace{1cm} 

\noindent \textit{Dedicated to Professor Rudi van Eldik on the occasion of his 70th Birthday.} 

\maketitle

\newpage

\section{Introduction}

The Configuration Interaction method (CI) is of great importance in quantum
mechanical calculations of the electron structure of atoms and molecules. It
is well known that the shortcomings of the CI method are due to the form of
the wave function which does not fulfill the electronic cusp condition \cite
{RW}:

\begin{equation}
\left( \frac 1\Psi \frac{\partial \Psi }{\partial r_{ij}}\right)
_{r_{ij}=0}=\frac 12.
\end{equation}
The CI wave function does not contain explicitly odd powers of the
interelectronic coordinate $r_{ij}$ (implicitly the CI wave function does
contain terms $r_{ij}^2$, $r_{ij}^4$, $\cdots $ $r_{ij}^{2n}$ formed by
combination of angular orbitals $p$, $d$, $f$ \cite{Drake,Sims02,Clary}).
But odd powers are the ones energetically important \cite{Weiss}. In the CI
wave function the piling up of higher and higher angular terms attempting to
represent the term $r_{ij}$ like in a Taylor expansion \cite{Weiss} shows
how important the details of the cusp are. The energy improvement when
increasing the quantum number $l$ follows an asymptotic formula proportional
to $(l+1/2)^{-4}$ \cite{Hill} in the case of helium atom, and in general for
a larger number of electrons \cite{Kutzelnigg,Schwartz62}. The nuclear cusp
condition is always fulfilled in the CI as in the Hartree-Fock (HF) wave
functions constructed with Slater orbitals:

\begin{equation}
\left( \frac 1\Psi \frac{\partial \Psi }{\partial r_i}\right) _{r_i=0}=-Z,
\end{equation}
$Z$ is the atomic charge, or the orbital exponent. The cusps (positive for
repulsion and negative for attraction) account for two-body correlation, but
not for three-body correlation.

These conditions are a result of the singularities of the Hamiltonian at $%
r_i=0$ and $r_{ij}=0$:

\begin{equation}
H=-\frac 12\sum_{i=1}^2\frac{\partial ^2}{\partial r_i^2}-\sum_{i=1}^2\frac
1{r_i}\frac \partial {\partial r_i}-\sum_{i=1}^2\frac 2{r_i}+\frac
1{r_{12}}-\frac 2{r_{12}}\frac \partial {\partial r_{12}}-\frac
12\sum_{i\neq j}^2\frac{r_i^2+r_{12}^2-r_j^2}{r_ir_{12}}\frac{\partial ^2}{%
\partial r_i\partial r_{12}}.
\end{equation}
As the exact wave function is obtained from the equation: $H\Psi /\Psi =E$, 
this equation leads to the exact energy only if the cusp conditions of Eqs.
(1,2) are fulfilled. The CI wave function is then not a formal solution of
the Schr\"{o}dinger equation and therefore the CI wave function converges
very slowly. Schwartz \cite{Schwartz} has demonstrated that the inclusion of
other terms than $r_{ij}$ into the wave function, like negative powers of $%
r_i$ \cite{Kinoshita}, fractionary powers of $r_i$ \cite{Thakkar} and
logarithmic terms $\ln (r_i+r_j)$ accelerates the convergence of the wave
function to the exact solution. Logarithmic terms are important to describe
the three-particle coalescence region \cite{Frankowski, Bartlett}. Nakatsuji 
\cite{Nakatsuji} has proposed the Iterative-Complement-Interaction (ICI)
method, which generates those terms which are necessary in the structure of
the wave function for extemely highly accurate calculations. Including the term $\ln
(r_1+r_2+\beta r_{12})$ in the wave function an energy of about $40$ decimal digits 
accuracy was obtained, showing that logarithmic terms are necessary \cite{Nakashima}. 
An improvement to the logarithmic terms provides the exponential integral (Ei), which does not show an unphysical   
node along the radial coordinate and it is a continuously decaying function. Including $Ei[-(r_1 +r_2)]$ terms into the wave function,   
an accuracy over 40 decimal digits was achieved for the non-relativistic energy of the $1^1S$ state of the 
helium atom$^1$\footnotetext[1]{Along this paper we are citing  
the ground state energy of helium atom calculated by the ICI method \cite{Kurokawa}, worlwide best value: 
-2.9037 2437 7034 1195 9831 1159 2451 9440 4446 6969 924 865 $E_h$ as exact 
value for chemical and physical purposes. For convenience we write down in 
the tables only 20 decimal digits.}. Similar calculations have been also carried out for  
two-electron ions \cite{Nakashima} and excited states \cite{He-excit}. Such
accurate calculations are restricted to the case of two-electron systems.

Other shortcomings of the CI wave function result from the functions used as
orbitals. If the functions form a complete set it is obvious that the exact
solution can be expressed in terms of these functions. Some types of orbitals
form a complete set (or overcomplete in case of double basis) like Slater
orbitals and Sturmians. If the set is not complete the wave function needs
some states from the continuum. The importance of the continuum functions in
representing the ground state of helium atom has been pointed out by many
authors \cite{SL,Green}. If one carries out calculations to the ultimative
limit it is immaterial which set is used. Therefore an infinite expansion of
CI configurations would be the exact wave function. The reason why the CI
wave function gives good results is the consecutive inclusion of angular
functions which represent $r_{ij}$.

It is then clear that the explicit correlated wave functions including the
interelectronic coordinate $r_{ij}$ proposed by Hylleraas \cite{Hylleraas}
are an alternative to the CI wave function. Sims and Hagstrom \cite
{Sims71,Sims71JCP} introduced the Hylleraas-Configuration Interaction
(Hy-CI) wave function which combines the use of higher angular momentum
orbitals as CI does with the inclusion of the interelectronic distance into
the wave function as the Hylleraas-type wave functions. The first terms of a
Hylleraas-CI wave function are CI terms.

Helium atom has been subject of numerous investigations. Its electronic
structure has been determined to the greatest accuracy known in quantum
chemistry. Our interest is to test our integral subroutines and computational 
techniques as the spin, antisymmetrization,
construction of symmetry adapted configurations, the usage of the
Hamiltonian in Hylleraas coordinates in the case of non-relativistic energy calculations on the
ground state of the helium atom, and so to gain experience towards investigating 
larger systems.

This work will be thoroughly referred and compared first, to the early
helium calculation of Weiss \cite{Weiss} for which there is enough
computational data, and second, to the highly accurate more modern
Configuration Interaction and Hylleraas-Configuration Interaction (Hy-CI)
calculations of Sims and Hagstrom \cite{Sims02}. The CI calculation of Weiss
is taken as starting point to check our program code. The CI and Hy-CI
calculations of Sims and Hagstrom have been taken as the
s,p,d-limits for comparison. Finally we have reproduced Hy-CI\ calculations of Sims and
Hagstrom \cite{comm}.

Novel in this work is the use of a different technique to evaluate the
kinetic energy Hy-CI integrals which we have used to reproduce the energy
values obtained from Sims and Hagstrom.

One of the purposes of this paper is to provide to the reader and future researchers 
in the field with detailed and accurate energy values, which can be used to write and 
specially to throroughly test computer programs for two-electron systems 
in various methodologies.

\section{The Hylleraas-CI wave function}

The Hy-CI wave function \cite{Sims71JCP} for the $1^1S$ ground state of
helium atom is written: 
\begin{equation}
\Psi =\sum_{k=1}^NC_k\Phi _k,\qquad \Phi _k=\hat{O}(\hat L^2)\hat{\mathcal{A}}%
\psi _k\chi ,\qquad \chi =(\alpha \beta -\beta \alpha ),
\end{equation}
\noindent where $\Phi _k$ are symmetry adapted configurations, $N$ is the
number of configurations and the constants $C_k$ are determined
variationally. The operator $\hat{O}(\hat L^2)$ projects over the proper spatial
space, so that every configuration is eigenfunction of $\hat{L}^2$. $\hat{%
\mathcal{A}}$ is the 2-particle antisymmetrization operator, and $\chi $ is
the spin eigenfunction. The Hartree products $\psi _k$, are products of
orbitals of the required symmetry multiplied by the interelectronic
coordinate:

\begin{equation}
\psi _k=\phi _{ik}(1)\phi _{ik}(2)r_{12}^\nu ,\qquad \nu =0,1
\end{equation}

For $\nu =0$ the Hylleraas-CI wave function reduces to the CI wave function.
The power $\nu =1$ does not represent any restriction. As discussed in Refs. 
\cite{Drake,RuizB,SimsH2}, we know that even powers of the interelectronic
coordinate are equivalent to products of $p$-, $d$-, $\cdots $ type
orbitals, for instance: 
\begin{equation}
r_{12}^2\equiv p(1)p(2).
\end{equation}
Furthermore, higher odd powers can be expresses as $r_{12}r_{12}^{2n}$. It
can be demonstrated \cite{RuizB} that $r_{12}$ is an infinite expansion of
angular orbitals: 
\begin{equation}
s(1)s(2)r_{12}\equiv s(1)s(2)+p(1)p(2)+d(1)d(2)+f(1)f(2)+\ldots
\end{equation}
In the case of two-electrons systems Sims and Hagstrom \cite{Sims02} have
shown that the strictly Hylleraas (Hy) and Hy-CI wave functions are equivalent.

\section{The Hamiltonian}

The non-relativistic Hamiltonian in the infinite nuclear mass approximation can be 
effectively written in Hylleraas coordinates for a Hy-CI wave function as$^2$ \footnotetext[2]{
As the Hy-CI wave function consists on only one $r_{ij}$ per configuration
and treating now helium atom with only two electrons, the terms of the
Hamiltonian including $\frac{\partial ^2} {\partial r_{12}^2}$ and $\frac{%
\partial ^2} {\partial r_{12}\partial r_{12}}$ vanish.} \cite{Ruiz1}:

\begin{multline}
\hat{H}=-\frac 12\sum_{i=1}^2\frac{\partial ^2}{\partial r_i^2}%
-\sum_{i=1}^2\frac 1{r_i}\frac \partial {\partial r_i}-\sum_{i=1}^2\frac
2{r_i}+\frac 1{r_{12}}-\frac 2{r_{12}}\frac \partial {\partial r_{12}}-\frac
12\sum_{i\neq j}^2\frac{r_i^2+r_{12}^2-r_j^2}{r_ir_{12}}\frac{\partial ^2}{%
\partial r_i\partial r_{12}} \\
-\frac 12\sum_{i=1}^2\frac 1{r_i^2}\frac{\partial ^2}{\partial \theta _i^2}%
-\frac 12\sum_{i=1}^2\frac 1{r_i^2\sin ^2{\theta _i}}\frac{\partial ^2}{%
\partial \varphi _i^2}-\frac 12\sum_{i=1}^2\frac{\cot {\theta _i}}{r_i^2}%
\frac \partial {\partial \theta _i} \\
-\sum_{i\neq j}^2\left( \frac{r_j}{r_ir_{12}}\frac{\cos {\theta _j}}{\sin {%
\theta _i}}+\frac 12\cot {\theta _i}\frac{r_{12}^2-r_i^2-r_j^2}{r_i^2r_{12}}%
\right) \frac{\partial ^2}{\partial \theta _i\partial r_{12}}-\sum_{i\neq
j}^2\frac{r_j}{r_ir_{12}}\frac{\sin {\theta _j}}{\sin {\theta _i}}\sin {%
(\varphi _i-\varphi _j)}\frac{\partial ^2}{\partial \varphi _i\partial r_{12}%
}\ .
\end{multline}
Using the Hamiltonian in this form it is easy to derive the involved integrals.
The angular momentum operator can be extracted and its eigenvalue equation
used:

\begin{equation}
\sum_{i=1}^2\frac 1{r_i^2}\hat{L}_i^2=-\frac 12\sum_{i=1}^2\frac 1{r_i^2}%
\frac{\partial ^2}{\partial \theta _i^2}-\frac 12\sum_{i=1}^2\frac
1{r_i^2\sin ^2{\theta _i}}\frac{\partial ^2}{\partial \varphi _i^2}-\frac
12\sum_{i=1}^2\frac{\cot {\theta _i}}{r_i^2}\frac \partial {\partial \theta
_i},
\end{equation}

\begin{equation}
L_i^2\phi _i=l_i(l_i+1)\phi _i,
\end{equation}
with $l_i$ the angular quantum number of the orbital $\phi _i$.

From the variational principle one obtains the matrix eigenvalue problem:

\[
(\mathbf{H}-E\mathbf{\Delta })\mathbf{C}=\mathbf{0,} 
\]
where the matrix elements are:

\begin{equation}
H_{kl}=\int \Phi _kH\Phi _ld\tau  ,\qquad  \Delta _{kl}=\int \Phi _k\Phi _ld\tau .
\end{equation}

\section{The use of real Slater orbitals}

We have constructed a set of real $s$-, $p$-, $d$-Slater orbitals which are
orthogonal and unnormalized. The exponents are considered as adjustable
parameters. They are defined as:
\begin{eqnarray}
&&ns=r^{n-1}e^{-\alpha r}  \nonumber \\
&&np_z=r^{n-1}e^{-\beta r}\cos (\theta )  \nonumber \\
&&np_x=r^{n-1}e^{-\beta r}\sin (\theta )\cos (\varphi )  \nonumber \\
&&np_y=r^{n-1}e^{-\beta r}\sin (\theta )\sin (\varphi )  \nonumber \\
&&nd_{z^2}=r^{n-1}e^{-\gamma r}\left( \frac 32\cos ^2(\theta )-\frac
12\right)  \nonumber \\
&&nd_{xz}=r^{n-1}e^{-\gamma r}\sin (\theta )\cos (\theta )\cos (\varphi ) 
\nonumber \\
&&nd_{yz}=r^{n-1}e^{-\gamma r}\sin (\theta )\cos (\theta )\sin (\varphi ) 
\nonumber \\
&&nd_{xy}=r^{n-1}e^{-\gamma r}\sin ^2(\theta )\cos (\varphi )\sin (\varphi )
\nonumber \\
&&nd_{{x^2}-{y^2}}=r^{n-1}e^{-\gamma r}\sin ^2(\theta )(\cos ^2(\varphi
)-\sin ^2(\varphi ))
\end{eqnarray}
The orbitals are eigenfunctions of $\hat{L}^2$, but they are not
eigenfunctions of $\hat{L}_z$.

\subsection{CI\ Integrals over real Slater orbitals}

The evaluation of the resulting matrix elements Eq. (11) for a CI wave
function leads to products of one- and two-electron integrals. These
integrals can be evaluated as shown in \cite{Ruiz2}. After performing the angular 
integration we write here compact
expression of all non-vanishing two-electron integrals$^3$ \footnotetext[3]{%
The symmetry of the two-electron integrals should be taken into account. We
write here only one case for each symmetry.} over $s$- and $p$-orbitals:

\begin{eqnarray}
&&\langle s(1)s(2)\frac 1{r_{12}}s(1)s(2)\rangle =(4\pi )^2\biggl< \frac
1{g_{12}}\biggr> ,  \nonumber \\
&&\langle s(1)p_i(2)\frac 1{r_{12}}s(1)p_i(2)\rangle =\frac{(4\pi )^2}3%
\biggl< \frac 1{g_{12}}\biggr> ,\qquad i=x,y,z  \nonumber \\
&&\langle p_i(1)s(2)\frac 1{r_{12}}s(1)p_i(2)\rangle =\frac{(4\pi )^2}9%
\biggl< \frac{s_{12}}{g_{12}^2}\biggr> ,\qquad i=x,y,z  \nonumber \\
&&\langle p_i(1)p_i(2)\frac 1{r_{12}}p_i(1)p_i(2)\rangle =(4\pi )^2\biggl< %
\frac 19\frac 1{g_{12}}+\frac 4{225}\frac{s_{12}^2}{g_{12}^3}\biggr> ,\qquad
i=x,y,z  \nonumber \\
&&\langle p_i(1)p_j(2)\frac 1{r_{12}}p_i(1)p_j(2)\rangle =(4\pi )^2\biggl<%
\frac 19\frac 1{g_{12}}-\frac 2{225}\frac{s_{12}^2}{g_{12}^3}\biggr> ,\qquad
i\ne j,\qquad i,j=x,y,z  \nonumber \\
&&\langle p_i(1)p_j(2)\frac 1{r_{12}}p_j(1)p_i(2)\rangle =(4\pi )^2\biggl<%
\frac 1{75}\frac{s_{12}^2}{g_{12}^3}\biggr> ,\qquad i\ne j,\qquad i,j=x,y,z
\end{eqnarray}
where $s_{12}$ is the smallest of $r_1$ and $r_2,$ and $g_{12}$ the largest
of $r_1$ and $r_2$. The integrals on the right hand side are radial integrals, which 
include only the radial part of the orbitals of Eq. (12). 
These integrals are expanded in terms
of auxiliary integrals \cite{Ruiz2}. The non-vanishing integrals over $%
d$-orbitals can be classified into three groups. The integrals over $s$- and $d$-orbitals:

\begin{eqnarray}
&&\langle s(1)d_{zz}(2)\frac{1}{r_{12}}s(1)d_{zz}(2)\rangle =\frac{(4\pi
)^{2}}{25}\biggl<\frac{1}{g_{12}}\biggr>, \nonumber \\
&&\langle s(1)d_{zz}(2)\frac{1}{r_{12}}d_{zz}(1)s(2)\rangle =\frac{(4\pi
)^{2}}{25}\biggl<\frac{s_{12}^2}{g_{12}^{3}}\biggr>, \nonumber \\
&&\langle s(1)d_{iz}(2)\frac{1}{r_{12}}s(1)d_{iz}(2)\rangle =\frac{(4\pi
)^{2}}{75}\biggl<\frac{s_{12}^2}{g_{12}^3}\biggr>,\qquad i=x,y \nonumber \\
&&\langle s(1)d_{iz}(2)\frac{1}{r_{12}}d_{iz}(1)s(2)\rangle =\frac{(4\pi
)^{2}}{75}\biggl<\frac{s_{12}^2}{g_{12}^{3}}\biggr>,\qquad i=x,y 
\end{eqnarray}

integrals over $p$- and $d$-orbitals: 
\begin{eqnarray}
&&\langle p_{x}(1)d_{xz}(2)\frac{1}{r_{12}}p_y(1) d_{yz}(2)\rangle =\frac{%
(4\pi )^{2}}{525}\biggl<\frac{s_{12}^2}{g_{12}^3}\biggr>, \nonumber \\
&&\langle p_{x}(1)d_{xz}(2)\frac{1}{r_{12}}d_{yz}(1)p_y(2)\rangle =\frac{%
(4\pi )^{2}}{735}\biggl<\frac{s_{12}^3}{g_{12}^4}\biggr>, \nonumber \\
&&\langle p_{i}(1)d_{iz}(2)\frac{1}{r_{12}}p_z(1) d_{zz}(2)\rangle =\frac{%
(4\pi )^{2}}{525}\biggl<\frac{s_{12}^2}{g_{12}^3}\biggr>,\qquad i=x,y \nonumber \\
&&\langle p_{i}(1)d_{iz}(2)\frac{1}{r_{12}}d_{zz}(1)p_z(2)\rangle =(4\pi)^2%
\biggl<-\frac{1} {225}\frac{s_{12}}{g_{12}^2}+\frac{4}{1225}\frac{s_{12}^{3}%
}{g_{12}^{4}}\biggr>,\qquad i=x,y \nonumber \\
&&\langle p_{z}(1)d_{zz}(2)\frac{1}{r_{12}}p_{z}(1)d_{zz}(2)\rangle =(4\pi
)^{2}\biggl<\frac{1}{25}\frac{1}{g_{12}}-\frac{4}{525}\frac{s_{12}^{2}}{%
g_{12}^{3}}\biggr>,\qquad \nonumber \\
&&\langle p_{z}(1)d_{zz}(2)\frac{1}{r_{12}}d_{zz}(1)p_{z}(2)\rangle =(4\pi
)^{2}\biggl<\frac{1}{225}\frac{s_{12}}{g_{12}^2}-\frac{9}{1225}\frac{%
s_{12}^{3}}{g_{12}^{4}}\biggr>, \nonumber \\
&&\langle p_{i}(1)d_{iz}(2)\frac{1}{r_{12}}p_{i}(1)d_{iz}(2)\rangle =(4\pi
)^{2}\biggl<\frac{1}{45}\frac{1}{g_{12}}+\frac{2}{1575}\frac{s_{12}^{2}}{%
g_{12}^{3}}\biggr>,\qquad i=x,y \nonumber \\
&&\langle p_{i}(1)d_{iz}(2)\frac{1}{r_{12}}d_{iz}(1)p_{i}(2)\rangle =(4\pi
)^{2}\biggl<\frac{1}{225}\frac{s_{12}}{g_{12}^2}+\frac{8}{3675}\frac{%
s_{12}^{3}}{g_{12}^{4}}\biggr>,\qquad i=x,y 
\end{eqnarray}
and integrals over $s$-, $p$-, and $d$-orbitals: 
\begin{eqnarray}
&&\langle d_{iz}(1)s(2)\frac{1}{r_{12}}p_i(1)p_z(2)\rangle =\frac{(4\pi )^{2}%
}{45}\biggl<\frac{s_{12}}{g_{12}^2}\biggr>, \qquad i=x,y \nonumber \\
&&\langle p_i(1) p_{z}(2)\frac{1}{r_{12}}s(1)d_{iz}(2)\rangle =\frac{(4\pi
)^{2}}{45}\biggl<\frac{s_{12}}{g_{12}^2}\biggr>, \qquad i=x,y \nonumber \\
&&\langle d_{iz}(1)p_i(2)\frac{1}{r_{12}}s(1)p_z(2)\rangle =\frac{(4\pi )^{2}%
}{75}\biggl<\frac{s_{12}^2}{g_{12}^3}\biggr>. \qquad i=x,y 
\end{eqnarray}

These expressions have been obtained with the help of the algebraic computer
program Maple \cite{Maple}.

\subsection{Construction of symmetry adapted configurations}

The ground state configuration of helium atom in the spectroscopic notation
is of the type $ss$ and has $S$ symmetry. We can also construct configurations
of $S$ symmetry using $p$-, $d$- orbitals. These configurations are:

\begin{eqnarray}
\psi _s &=&ns(1)ns(2)=ss=s^2, \nonumber \\
\psi _p &=&np(1)np(2)=pp=p^2, \nonumber \\
\psi _p &=&nd(1)nd(2)=dd=d^2,
\end{eqnarray}
with

\begin{equation}
\psi _p=pp=p_x(1)p_x(2)+p_y(1)p_y(2)+p_z(1)p_z(2)=p_xp_x+p_yp_y+p_zp_z.
\end{equation}
The basis functions $p_yp_y$ and $p_zp_z$ may not contribute energetically
as much as $p_xp_x$ but they are necessary to have the proper symmetry, so
that the wave function is eigenfunction of $L^2$. Therefore a matrix element
between two of these configurations has to be calculated as:

\begin{equation}
\left\langle pp\left| \hat{H}\right| pp\right\rangle =\left\langle
p_xp_x+p_yp_y+p_zp_z\left| H\right| p_xp_x+p_yp_y+p_zp_z\right\rangle .
\end{equation}
One has to note that using symmetry adapted functions or configurations
(SAF) and non-SAF the dimensions of the $H$-matrices are different, whereas
a proof of the correctness of the calculation is that the diagonalization of
the $H$-matrices in both cases should lead to the exactly same energy
results \cite{comm}. We have used the notation $pp$, but different exponents
or powers may be used, $pp^{\prime }.$

In the case of $d$-orbitals:

\begin{equation}
\psi
_d=dd=d_{z^2}d_{z^2}+d_{xz}d_{xz}+d_{yz}d_{yz}+d_{xy}d_{xy}+d_{x^2-y^2}d_{x^2-y^2}.
\end{equation}

\section{The use of complex Slater orbitals}

The complex Slater orbitals with quantum numbers $n,m$ and $l$ are defined
by an unnormalized radial part and an angular orthonormal part which is a
spherical harmonic: 
\begin{eqnarray}
\phi ^{*}(\mathbf{r}) &=&r^{n-1}e^{-\alpha r}Y_l^{m*}(\theta ,\phi ), 
\nonumber \\
\phi ^{\prime }(\mathbf{r}) &=&r^{n^{\prime }-1}e^{-\alpha ^{\prime
}r}Y_{l^{\prime }}^{m^{\prime }}(\theta ,\phi ).
\end{eqnarray}
The spherical harmonics in Condon and Shortley phases \cite[p. 52]{Condon}
are given by: 
\begin{equation}
Y_l^m(\theta ,\phi )=(-1)^m\left[ \frac{2l+1}{4\pi }\frac{(l-m)!}{(l+m)!}%
\right] ^{1/2}P_l^m(\cos {\theta })e^{im\phi },
\end{equation}
where $P_l^m(\cos {\theta })$ are the associated Legendre functions. The
spherical harmonics and associated Legendre functions used along this work
are written explicitly in \cite[p. 14]{Stevenson}.

The charge distributions are: 
\begin{equation}
\Omega _{N,L,M}(\mathbf{r})=\phi ^{*}(\mathbf{r})\phi ^{\prime }(\mathbf{r}
)=\sum_{L=|l-l^{\prime }|}^{l+l^{\prime }}{}(2L+1)^{1/2}C^L(l^{\prime
},m^{\prime };l,m)r^{N-1}e^{-\omega r}Y_L^M(\theta ,\phi ),
\end{equation}
where $N=n+n^{\prime }-1$, $M=m^{\prime }-m$ and the exponents $\omega
=\alpha +\alpha ^{\prime }$. $N,L,M$ are the quantum numbers of the charge
distributions. $C^L(l^{\prime },m^{\prime };l,m)$ are the Condon and
Shortley coefficients \cite[p. 52]{Condon}.

\subsection{Integrals over complex Slater orbitals}

The two-electron integrals appearing in the Hy-CI method are then defined: 
\begin{equation}
I(N_1,N_2;\omega _1,\omega _2;\nu )_{l_1,l_1^{\prime },l_2,l_2^{\prime
}}^{m_1,m_1^{\prime },m_2,m_2^{\prime }}=\int \Omega _{N_1,L_1,M_1}(\mathbf{r%
}_1)\Omega _{N_2,L_2,M_2}(\mathbf{r}_2)r_{12}^\nu d\mathbf{r}_1d\mathbf{r}_2,
\end{equation}
with the cases $\nu =-1,0,1,2$. After angular integration the two-electron
integrals are: 
\begin{multline}
I(N_1,N_2;\omega _1,\omega _2;\nu )_{l_1,l_1^{\prime },l_2,l_2^{\prime
}}^{m_1,m_1^{\prime },m_2,m_2^{\prime }}=(-1)^{M_1}\delta
(M_1+M_2,0)\sum_{L_1=|l_1-l_1^{\prime }|}^{l_1+l_1^{\prime
}}\sum_{L_2=|l_2-l_2^{\prime }|}^{l_2+l_2^{\prime }}\delta (L_1,L_2) \\
\times \prod_{i=1}^2(2L_i+1)^{1/2}C^{L_i}(l_i^{\prime },m_i^{\prime
};l_i,m_i)I(N_1,N_2;\omega _1,\omega _2;\nu ;L_2), 
\end{multline}
where $I(N_1,N_2;\omega _1,\omega _2;\nu ;L_2)$ are the basic radial
two-electron integrals \cite{Ruiz2e}. The occurring basic two-electron integrals in the
calculations of helium atom are:

\begin{equation}
I(N_1,N_2;\omega _1,\omega _2;0;L)=\delta (L,0)A(N_1+1,\omega
_1)A(N_2+1,\omega _2), 
\end{equation}

\begin{multline}
I(N_1,N_2;\omega _1,\omega _2;1;L)=\frac 1{(2L+1)} \\
\times \left[ -\frac 1{(2L-1)}\left[ V(N_1+L+1,N_2-L+2;\omega _1,\omega
_2)+V(N_2+L+1,N_1-L+2;\omega _2,\omega _1)\right] +\right. \\
+\left. \frac 1{(2L+3)}\left[ V(N_1+L+3,N_2-L;\omega _1,\omega
_2)+V(N_2+L+3,N_1-L;\omega _2,\omega _1)\right] \right], 
\end{multline}
\begin{equation}
I(N_1,N_2;\omega _1,\omega _2;-1;L)=\frac 1{(2L+1)}\left[
V(N_1+L+1,N_2-L;\omega _1,\omega _2)+V(N_2+L+1,N_1-L;\omega _2,\omega
_1)\right] ,
\end{equation}
\begin{multline}
I(N_1,N_2;\omega _1,\omega _2;2;L)=\delta (L,0)\left[ A(N_1+3,\omega
_1)\right. A(N_2+1,\omega _2)+\left| A(N_2+3,\omega _2)A(N_1+1,\omega
_1)\right] \\
-\frac 23\delta (L,1)A(N_1+2,\omega _1)A(N_2+2,\omega _2), 
\end{multline}
$A(n,\alpha )$ and $V(k,l;\alpha ,\beta )$ are auxiliary two-electron
integrals defined in Ref. \cite{Ruiz3e}.

\subsection{Hylleraas-CI two-electron kinetic energy integrals}

For any atomic number $N$ $\geq $ $3$ the kinetic energy integrals
are of two- and three-electron type. The three-electron kinetic energy
integrals have been evaluated in Refs. \cite{Sims3e,Ruiz3e}. For $N=2$ the
integrals are only of two-electron type. Sims and Hagstrom \cite
{Sims3e} evaluated the two-electron kinetic energy integrals using the transformation of
Kolos and Roothaan \cite{KR}, which partially avoids the differentiation with
respect to $r_{ij}$ terms appearing on the right hand side of the matrix
elements. In a previous work \cite{Ruiz2e} we have evaluated the two-electron kinetic energy 
integrals performing the derivatives directly over $r_{ij}$
and using the Hamiltonian written in polar and interelectronic coordinates \cite{Ruiz1}. In 
this work we have used these expressions to calculate the ground state of helium atom and 
therefore proved the correctness of the expressions of Ref. \cite{Ruiz2e}. 

Further the alternative method of the Kolos and Roothaan transformation has been also 
implemented in the computer code. Both methods lead computationally exactly to the 
same results, confirming the formulas of the Appendix B \cite{Ruiz2e}. 
Moreover the calculations of the ground state of helium atom have proven to be 
computationally faster when using the Kolos and Roothaan algorithm. 
Nevertheless it is important to test the direct method 
of evaluation, because this is the one which has to be used for the three-electron kinetic energy 
integrals when using the Hamiltonian in Hylleraas coordinates, since in the three-electron case 
the Kolos and Roothaan transformation cannot be used \cite{Ruiz3ek}.

\subsection{Construction of symmetry adapted configurations CI and Hy-CI}

The configurations $ss$, $pp$, $dd$, $ff$ of $S$ symmetry using complex Slater
orbitals are constructed:

\begin{equation}
\psi
_p=pp=p_0(1)p_0(2)-p_1(1)p_{-1}(2)-p_{-1}(1)p_1(2)=p_0p_0-p_1p_{-1}-p_{-1}p_1
\end{equation}
the basis functions $p_1p_{-1}$ and $p_{-1}p_1$ are energetically
degenerated. A matrix element between two of these configurations has to be
calculated as:

\begin{equation}
\left\langle pp\left| \hat{H}\right| pp\right\rangle =\left\langle
p_0p_0-p_1p_{-1}-p_{-1}p_1\left|\hat H\right|
p_0p_0-p_1p_{-1}-p_{-1}p_1\right\rangle
\end{equation}

The configuration $dd$ of $S$ symmetry is constructed:

\begin{equation}
\psi _d=dd=d_0d_0-d_1d_{-1}-d_{-1}d_1+d_2d_{-2}+d_{-2}d_2
\end{equation}
the basis functions $d_1d_{-1}$ and $d_{-1}d_1$ are degenerated, also $%
d_2d_{-2}$ and $d_{-2}d_2$.

The $ff$ configuration of S symmetry is: 

\begin{equation}
\psi _f=ff=f_0f_0-f_1f_{-1}-f_{-1}f_1+f_2f_{-2}+f_{-2}f_2-f_3f_{-3}+f_{-3}f_3
\end{equation}
the basis functions $f_1f_{-1}$ and $f_{-1}f_1$ are degenerated, also $%
f_2f_{-2}$ and $f_{-2}f_2$ and $f_3f_{-3}$ and $f_{-3}f_3$.

These rules are also fulfilled for configurations constructed with a pair of orbitals with 
different powers or exponents.

The Hy-CI configurations are then constructed by multiplying a Hartree product by the 
interelectronic coordinate (note 
that a configuration is already antisymmetrized, whereas in the Hy-CI wave function the 
antisymmetrization operator acts also on the $r_{ij}$ factor). The $r_{ij}$ factor has radial 
symmetry and therefore it does not affect the total angular momentum $L$ of the 
configuration. Therefore the Hy-CI SAF are constructed in the same way than the CI ones.

\section{Calculations}

\subsection{CI Calculations}

In order to test the computer code the $s,p,d$-calculations of Weiss (Table II) \cite{Weiss} 
were reproduced. The recalculated values are shown with more decimal digits in
Table I. The CI codes using real and complex orbitals led to the same energy
results. Both codes use quadruple precision (QP), about $30$ decimal digits in
our computer$^4$ \footnotetext[4]{All the calculations of this paper have been carried out 
in a single computer without parallelization. The computer is an Intel Xeon E3-1240 
V2 3.4 GHz machine with 4 processors. The compiler used is the Intel Fortran Composer XE  
2013 for Linux. The program is compiled automatically to QP with the command \textit{ifort -r16 -i8 -o program.x program.f}.}. 
The advantage of complex orbitals is that general algorithms
can be developed and applied, as the angular integration, and general
integral subroutines. 

We have tried first the full optimization of the CI wave function. This technique had to 
be abandoned because it become impracticable. 
In Table II we show a set of optimized exponents. They have been optimized
in the field of about $10$ configurations. We have also tried single and double, and half basis sets, these last
were built up by multipliying the exponents of the single basis by $2$ or
dividing by $2$. The double basis could be used until the exponent with
value $20.0$, further the quadruple precision was not enough. The half basis
could be used only for some functions because linear dependence appeared
(too similar exponents). 

The CI calculations of Table III and the Hy-CI ones of Table V have been performed using different exponents 
for different orbitals, beeing these exponents optimized, while 
the calculations of Tables VI-VIII have been done using a single orbital exponent for both electrons and 
at the same time for all configurations. The optimization of the single orbital exponent is done then in the field 
of all configurations. 
We have observed that, in the practice, the energy result for  
long wave function expansions depends very little on the orbital exponents. 
Nevertheless we employ in this work different exponents in the first calculations, because this is 
very helpful for testing purposes. 
The first $15$ terms of the CI wave function in Table III are the same than
the ones used by Weiss, they gave a good energy result. We also used the
first $10$ $p$-functions of Weiss. Afterwards new terms are added
systematically. The configurations are ordered by orbital type showing the pattern of convergence and therefore the existence of 
the $s$-limit, $p$-limit, $d$-limit, \ldots in the CI calculations. The precise energy contribution per configuration is shown to facilitate 
the testing of computer programs. We show here the convergence of the $101$-term CI wave
function as an example. 

The convergence pattern of the CI wave function was
very discouraging. In Table IV we show our best results grouped as $s$-limit, $p$-limit and $d$-limit obtained using QP. 
They agree with the results of Sims and Hagstrom who
used real*$24$ arithmetic and with the results of Bromley and Mitroy \cite
{Bromley} using Laguerre functions. A $668$ term CI calculation leads to 
-2.9027 6626 64 a.u. which is about $1.0$ millihartree accurate. The best
calculation up to date and the largest expansion is the one of Sims and
Hagstrom which is about $4$ microhartree above the exact value. 

These facts confirm the foreseeing from many authors. i.e. Carrol, Silverstone and
Metzger \cite{Silverstone} and Kutzelnigg and Morgan \cite{Kutzelnigg} pointed out:
''For a two electron atom using conventional CI to achieve microhartree
accuracy (they mentioned an accuracy of $20$ microhartrees) requires
enormous labor and sophisticacion, and indicates that further improvement in
accuracy by brutal force CI would be extremely difficult, if not virtually
impossible because of numerical linear dependence problems.''

\subsection{CI-R12 Calculations}

The R12 method can be considered as a restricted Hy-CI method, with the restriction, only the 
interelectronic distance $r_{12}$ is included into the wave function. Also in this method 
not only two-electron integrals occur but also three- and four-electron ones. These last 
are solved approximately using the so-called resolution of identity. Therefore the R12 method is not strictly 
variational, nevertheless it is mostly used in numerous applications, especially molecular calculations. 
The R12 wave function does not fulfill the nuclear Kato cusp condition, since Gaussian orbitals 
are used. For details about this method, see Ref. \cite{Klopper}. 

With our Hy-CI computer code we can also perform calculations with the CI-R12 method, which 
consists in the strictly variational R12 method counterpart solving all the occurring integrals 
analytically and using Slater orbitals. In the case of the helium atom the CI-R12 and the Hy-CI 
wave functions are equivalent, and the calculations led to the same result, as shown in Table IX. 

Usually in the R12 method only correlated configurations which are constructed with 
the ground state configuration are included. 
The energy value according to our program is -2.9034 9813 3768 $E_h$, 
which is 0.2 millihartree accurate and the computational time was 138 seconds. 
If as in the Hy-CI wave function also excited correlated configurations other than the 
ground state one would be considered, then the energy value will be improved to higher accuracy. 
In case of the He atom these calculations are equivalent to the Hy-CI ones of Table IX with 
orbitals of maximum L=1,2,3.

\subsection{Hy-CI Calculations}

We have taken the wave function of Weiss and have added the same functions
containing $r_{12}$. The exponents were slighly optimized in a field of
about $10$ functions. The set of exponents of the $r_{12}$-configurations
has been taken different than the purely CI configurations, see Table I.
They are in general slightly larger. In Table V a $62$-term truncated
Hy-CI wave function is shown. One can observe the faster convergence of the
wave function leading to a result which is $0.5$ microhartree accurate. Note
that the computation time was $87$ seconds, although the program code has
not been refined to save time. 

If we use a single orbital exponent for electrons 1 and 2 and in every configuration, 
the amount of auxiliary integrals to be computed is drastically reduced. This accelerates  
the performance of the Hy-CI calculations. The exponent for the double occupied shell was obtained by 
repeated optimization of a shorter Hy-CI wave function. Extensive calculations using a single 
exponent are shown in Table VI using $s$-, $p$-, $d$-, and $f$-orbitals, 
\textit{i.e. }
allowed to obtain the energy -2.9037 2437 7024 3236 8211 $E_h$, a
result of picohartree accuracy in less than 2 minutes in our computer.

\subsection{Hy Calculations}

The developed integral equations are valid for any power value $\nu$ in $r_{ij}^{\nu}$. 
For $\nu \ge 2$ the wave function of Eqs. (3,4) is a Hylleraas-type wave function 
(strictly Hylleraas). The Hy-CI and Hy configurations in the case of the helium atom 
are equivalent and therefore the presence of both simultaneously in the 
wave function would lead to repetition and linear dependences.  
This has been tested in our calculations. The addition of a repeated or equivalent configuration 
does not add ($ <1.0\times 10^{-30}$ $E_h$) to the previous energy value. 
Linear dependence denotes that in a system of equations (the diagonalization procedure consist in 
solving the eigenvalue equation) one of the equations is not linear independent from the others, and therefore 
the equations have no one solution. In our work every configuration leads to a row or column in the 
system of equations. If one configuration is equivalent to another, the linear dependence problem occurs.   
 


In Table VII we show the Hy calculations using the same orbital exponents than in the 
Hy-CI calculations of Table VI. Correlated configurations including $p$-, $d$-, $f$-, 
$\ldots$ orbitals were dropped out because they were redundant. The convergence pattern of the 
truncated Hy expansion is faster than the one of the Hy-CI wave function, whereas in the Hy wave 
function precision problems appear earlier than in the Hy-CI calculations due to the simultaneous existence of small and huge 
values of the matrix elements. Then the eigenvalue equation has to be solved very accurately using higher 
precision arithmetic.  
The use of different orbital types in the 
Hy-CI expansion avoids early linear dependence problems as so as huge values of matrix elements 
and permits us to use very high orbitals N=20 
with QP in our program. The energy obtained was about 3 picohartree accurate, whereas the computational 
time of 2.3 minutes is slightly higher than for the Hy-CI wave function. For large wave function expansions it can be 
observed using our computer program that the Hy wave function needs larger computational times 
than the Hy-CI wave function. This could be due to the favourable orthogonalities between orbitals with different 
angular momentum in the Hy-CI wave function.  

If we order the configurations of the Hy wave function by the $\nu$ power, we obtain the 
truncated wave function expansion of Table VIII, where one can observe the weight of the configurations 
with low $\nu$ power, in particular the importance of the power $\nu=1$.

\section{Conclusions}

It has been shown that the CI wave function converges extremely slowly. A shorter
configuration expansion is needed by the Hy-CI wave function to obtain a
better result. Picoharteee accuracy has been achieved using a set of $s$-, $%
p$-, $d$-, and $f$-orbitals and $r_{12}$. The integral equations and CI and Hy-CI precise 
energy values presented in this paper can serve to program and thoroughly test computer codes 
for the two-electron systems. The presented CI and Hy-CI calculations show agreement with the
ones of Sims and Hagstrom and this confirms the correctness of the 
used Hamiltonian in Hylleraas coordinates and so as of the two-electron kinetic energy integrals 
methodologies.

Furher we have shown Hy calculations on the helium atom, which can be performed using the 
integral formulas for $\nu \ge 2$. We have compared the pattern of convernce and details of the 
calculations by these two methods. We have shown and discussed CI-R12 calculations in the case 
of helium atom. Exponential correlated wave functions have not been treated in this work, this 
could be done introducing the exponential correlation term into the equations and it would be 
interesting to see the pattern of convergence of the wave function.

\section*{Acknowledgments}

This paper is dedicated to Rudi van Eldik, Professor of Inorganic Chemistry at 
the Department of Chemistry and Pharmacy of our University. Electronic Structure 
calculations counts among his fields of interest and attention. 
This work treats about the noble gas helium, the smallest of the inorganic substances.  
The author is deeply indebted to Peter Otto for his teaching on electron structure 
computer programming, support and encouragement of this project.  
Finally the author would like to thank very much James S. Sims for interesting discussions
about the Hy-CI method and the helium atom, and also to James S. Sims and
Stanley A. Hagstrom for providing the values of some matrix elements to
check the program code.

\newpage


\begin{table}[tp]
\begin{center}
\caption{Repetition of Weiss calculations \cite{Weiss}.} 
\begin{tabular}{cccc}
\hline\hline
& Confs. & \qquad E ($E_h$) \qquad &  Virial \\ \hline
Weiss & 15 & -2.8789 5525 8604 & 1.999 999 \\ 
Weiss & 25 & -2.9003 9002 0339 & 1.958 324 \\ 
Weiss & 31 & -2.9025 8300 1014 & 2.000 020 \\ \hline\hline
\end{tabular}
\end{center}
\end{table}

\newpage


\begin{table}[tp]
\begin{center}
\caption{Extension of the Weiss wave function. Basis set used in the CI and in Hy-CI
calculations.}
\begin{tabular}{ccccc}
\hline\hline
Orbital & CI & Hy-CI & Orbital & CI \\ 
\hline
$1s$ & 1.48 & 1.9 & $9s$ & 9.7 \\ 
$2s$ & 1.48 & 1.9 & $10s$ & 10.7 \\ 
$1s^{\prime}$ & 3.7 & 3.9 & $11s$ & 11.7 \\ 
$2s^{\prime}$ & 3.7 & 3.4 & $12s$ & 12.7 \\ 
$3s$ & 3.7 & 3.0 & $13s$ & 13.7 \\ 
$4s$ & 4.7 & 5.4 & $14s$ & 14.7 \\ 
$5s$ & 5.7 & 4.4 & $15s$ & 15.7 \\ 
$6s$ & 6.7 & 5.5 & $9p$ & 5.7 \\ 
$7s$ & 7.7 & 8.4 & $10p$ & 6.0 \\ 
$8s$ & 8.7 & 8.8 & $11p$ & 6.3 \\ 
$2p$ & 2.7 & 3.0 & $12p$ & 6.7 \\ 
$3p$ & 2.7 & 2.0 & $13p$ & 7.0\\ 
$2p^{\prime}$ & 5.4 & 5.0 & $14p$ & 7.3 \\ 
$3p^{\prime}$ & 5.4 & 6.0 & $15p$ & 7.7 \\ 
$4p$ & 3.3 & 3.8 & $16p$ & 8.0 \\ 
$5p$ & 3.7 & 4.5 & $9d$ & 7.8 \\ 
$6p$ & 4.3 & 5.2 & $10d$ & 8.6 \\ 
$7p$ & 4.7 & 5.9 & $11d$ & 9.4  \\ 
$8p$ & 5.3 & 6.5 & $12d$ & 10.0 \\ 
$3d$ & 3.6 & 3.8 & $13d$ & 10.6 \\ 
$4d$ & 2.6 & 4.2 & $14d$ & 11.4 \\ 
$5d$ & 4.7 & 4.6 & $15d$ & 12.2 \\ 
$6d$ & 5.5 & 3.9 & $16d$ & 13.0 \\ 
$7d$ & 6.3 & 4.1 &  & \\ 
$8d$ & 7.1 & 4.6 &  & \\ 
\hline\hline
\end{tabular}
\end{center}
\end{table}

\newpage


\begin{table}[tp]
\begin{center}
\caption{Extension of the Weiss wave function: $s$-, $p$-, and $d$-CI calculations on the $1^1S$ ground state of helium
atom. A $101$ terms truncated wave function expansion using the exponents of Table II is presented. Configuration $45$ is  
the $s$-limit of this calculation, configuration $80$ the $p$-limit and configuration $101$ the $d$-limit. 
The calculation time was 403 seconds in our computer using QP.}
\begin{tabular}{rcccc}
\hline\hline
N & Configuration & \quad Energy ($E_h$)\qquad & \qquad Virial \qquad & \qquad 
Difference ($E_h$)\\ 
\hline
1 & $1s1s$ & -2.8046 0000 0000 & 2.28041 &  0.0$\;\;\;\;\;\;\;\;\;\;\;\;\;\;\;\;\;\;\;$ \\ 
2 & $1s2s$ & -2.8464 5649 2448 & 2.01266 & -0.041856492448 \\
3 & $2s2s$ & -2.8528 0522 2662 & 2.03876 & -0.006348730214 \\ 
4 & $1s1s^{\prime}$ & -2.8749 7028 4491 & 2.00205 & -0.022165061829 \\ 
5 & $2s1s^{\prime}$ & -2.8753 1452 7707 & 2.00131 & -0.000344243216 \\
6 & $1s^{\prime}1s^{\prime}$ & -2.8762 1971 7792 & 2.00040 & -0.000905190085 \\ 
7 & $1s2s^{\prime}$ & -2.8774 6546 4572 & 2.00105 & -0.001245746780 \\
8 & $2s2s^{\prime}$ & -2.8784 4308 9284 & 2.00007 & -0.000977624712 \\
9 & $1s^{\prime}2s^{\prime}$ & -2.8784 9065 1571 & 2.00005 & -0.000047562287 \\ 
10 & $2s^{\prime}2s^{\prime}$ & -2.8787 4806 7759 & 1.99986 & -0.000257416188 \\ 
11 & $1s3s$ & -2.8787 8660 6368 & 2.00004 & -0.000038538609 \\ 
12 & $2s3s$ & -2.8788 2609 4227 & 2.00009 & -0.000039487859 \\ 
13 & $1s^{\prime}3s$ & -2.8789 0233 9991 & 2.00003 & -0.000076245764 \\ 
14 & $2s^{\prime}3s$ & -2.8789 0276 7307 & 2.00002 & -0.000000427316 \\ 
15 & $3s3s$ & -2.8789 5525 8583 & 2.00000 & -0.000052491276 \\ 
16 & $1s4s$ & -2.8789 5642 2086 & 2.00000 & -0.000001163503 \\ 
17 & $2s4s$ & -2.8789 5663 3375 & 2.00000 & -0.000000211289 \\
18 & $3s4s$ & -2.8789 5751 8997 & 2.00001 & -0.000000885622 \\
19 & $4s4s$ & -2.8789 7440 1976 & 1.99999 & -0.000016882979 \\ 
20 & $1s5s$ & -2.8789 7575 8617 & 2.00000 & -0.000001356640 \\
21 & $2s5s$ & -2.8789 7576 2471 & 2.00000 & -0.000000003854 \\ 
22 & $3s5s$ & -2.8789 7720 3414 & 1.99999 & -0.000001440944 \\ 
23 & $4s5s$ & -2.8789 8084 9314 & 1.99999 & -0.000003645900 \\ 
24 & $5s5s$ & -2.8789 8702 4973 & 1.99998 & -0.000006175659 \\ 
\hline\hline
\end{tabular}
\end{center}
\end{table}

\newpage


\begin{table}[tp]
\begin{center}
\textbf{Table III} Continuation. \\[0pt]
\begin{tabular}{rcccc}
\hline\hline
N & Configuration & \quad Energy ($E_h$)\qquad & \qquad Virial \qquad & 
\qquad Difference ($E_h$) \\
\hline
25 & $1s6s$ & -2.8789 8769 6600 & 1.99999 & -0.000000671627 \\
26 & $2s6s$ & -2.8789 8772 2840 & 1.99999 & -0.000000026240 \\
27 & $3s6s$ & -2.8789 8826 1063 & 1.99999 & -0.000000538223 \\
28 & $4s6s$ & -2.8789 8868 8707 & 1.99998 & -0.000000427644 \\
29 & $5s6s$ & -2.8789 8975 9185 & 1.99998 & -0.000001070478 \\
30 & $6s6s$ & -2.8789 9115 9584 & 1.99998 & -0.000001400398 \\
31 & $1s7s$ & -2.8789 9262 2201 & 1.99999 & -0.000001462617 \\
32 & $2s7s$ & -2.8789 9262 9614 & 1.99999 & -0.000000007413 \\
33 & $3s7s$ & -2.8789 9327 4233 & 1.99999 & -0.000000644620 \\
34 & $4s7s$ & -2.8789 9335 5874 & 1.99999 & -0.000000081641 \\
35 & $5s7s$ & -2.8789 9346 3777 & 1.99999 & -0.000000107903 \\
36 & $6s7s$ & -2.8789 9377 7434 & 1.99999 & -0.000000313656 \\
37 & $7s7s$ & -2.8789 9419 5021 & 1.99999 & -0.000000417587 \\
38 & $1s8s$ & -2.8789 9565 2646 & 2.00000 & -0.000001457625 \\
39 & $2s8s$ & -2.8789 9565 4558 & 2.00000 & -0.000000001913 \\
40 & $3s8s$ & -2.8789 9600 4792 & 1.99999 & -0.000000350234 \\
41 & $4s8s$ & -2.8789 9606 3069 & 1.99999 & -0.000000058277 \\
42 & $5s8s$ & -2.8789 9608 4615 & 1.99999 & -0.000000021546 \\
43 & $6s8s$ & -2.8789 9611 7709 & 1.99999 & -0.000000033094 \\
44 & $7s8s$ & -2.8789 9624 1931 & 1.99999 & -0.000000124222 \\
45 & $8s8s$ & -2.8789 9639 3909 & 1.99999 & -0.000000151979 \\
46 & $2p2p$ & -2.8980 1066 3847 & 1.99720 & -0.019014269938 \\
47 & $2p3p$ & -2.8983 0353 5068 & 1.99927 & -0.000292871221 \\
48 & $3p3p$ & -2.8999 8445 0188 & 2.00053 & -0.001680915121 \\ 
49 & $2p2p^{\prime}$ & -2.9000 1939 5255 & 2.00040 & -0.000034945067 \\ 
50 & $3p2p^{\prime}$ & -2.9000 2572 8711 & 2.00030 & -0.000006333456 \\ 
51 & $2p^{\prime}2p^{\prime}$ & -2.9003 0029 7330 & 2.00012 & -0.000274568618 \\
52 & $2p3p^{\prime}$ & -2.9003 0993 6342 & 2.00014 & -0.000009639013 \\ 
53 & $3p3p^{\prime}$ & -2.9003 4614 9487 & 2.00001 & -0.000036213144 \\ 
\hline\hline
\end{tabular}
\end{center}
\end{table}

\newpage


\begin{table}[tp]
\begin{center}
\textbf{Table III} Continuation. \\[0pt]
\begin{tabular}{rcccc}
\hline\hline
N & Configuration & Energy ($E_h$) & Virial & Difference ($E_h$)\\
\hline
54 & $2p^{\prime}3p^{\prime}$ & -2.9003 4931 2787 & 2.00003 & -0.000003163300 \\
55 & $3p^{\prime}3p^{\prime}$ & -2.9004 2459 8409 & 2.00001 & -0.000075285622 \\
56 & $2p4p$ & -2.9004 3166 9931 & 2.00003 & -0.000007071522 \\
57 & $3p4p$ & -2.9004 3221 0413 & 2.00004 & -0.000000540482 \\
58 & $4p4p$ & -2.9004 3399 3582 & 2.00004 & -0.000001783169 \\ 
59 & $2p5p$ & -2.9004 3400 6808 & 2.00004 & -0.000000013225 \\ 
60 & $3p5p$ & -2.9004 3508 9064 & 2.00005 & -0.000001082256 \\ 
61 & $4p5p$ & -2.9004 4229 5990 & 2.00002 & -0.000007206927 \\ 
62 & $5p5p$ & -2.9004 4249 5331 & 2.00002 & -0.000000199340 \\
63 & $2p6p$ & -2.9004 4266 5968 & 2.00002 & -0.000000170637 \\
64 & $3p6p$ & -2.9004 4268 0235 & 2.00002 & -0.000000014267 \\ 
65 & $4p6p$ & -2.9004 4712 1025 & 2.00002 & -0.000004440789 \\ 
66 & $5p6p$ & -2.9004 5007 1856 & 2.00003 & -0.000002950831 \\ 
67 & $6p6p$ & -2.9004 5688 7282 & 2.00002 & -0.000006815427 \\
68 & $2p7p$ & -2.9004 5688 7957 & 2.00002 & -0.000000000675 \\ 
69 & $3p7p$ & -2.9004 5691 7483 & 2.00002 & -0.000000029525 \\
70 & $4p7p$ & -2.9004 6603 9818 & 2.00001 & -0.000009122335 \\
71 & $5p7p$ & -2.9004 6617 5403 & 2.00001 & -0.000000135584 \\ 
72 & $6p7p$ & -2.9004 6618 6237 & 2.00001 & -0.000000010834 \\ 
73 & $7p7p$ & -2.9004 7220 2637 & 2.00001 & -0.000006016400 \\ 
74 & $2p8p$ & -2.9004 7221 0819 & 2.00001 & -0.000000008181 \\ 
75 & $3p8p$ & -2.9004 7221 5238 & 2.00001 & -0.000000004419 \\ 
76 & $4p8p$ & -2.9004 7250 5447 & 2.00001 & -0.000000290209 \\ 
77 & $5p8p$ & -2.9004 7255 4026 & 2.00001 & -0.000000048579 \\
78 & $6p8p$ & -2.9004 7428 5314 & 2.00001 & -0.000001731287 \\ 
79 & $7p8p$ & -2.9004 7683 0335 & 2.00001 & -0.000002545022 \\ 
80 & $8p8p$ & -2.9004 7916 7253 & 2.00001 & -0.000002336918 \\ 
81 & $3d3d$ & -2.9022 4165 1033 & 2.00001 & -0.001762483780 \\                   
82 & $3d4d$ & -2.9022 4223 6322 & 1.99999 & -0.000000585289 \\                   
\hline\hline
\end{tabular}
\end{center}
\end{table}

\newpage


\begin{table}[tp]
\begin{center}
\textbf{Table III} Continuation. \\[0pt]
\begin{tabular}{rcccc}
\hline\hline
N & Configuration & \quad Energy ($E_h$) \qquad & \qquad virial \qquad & 
\qquad Difference ($E_h$) \\
\hline
83 & $4d4d$ & -2.9024 0837 2063 & 2.00023 & -0.000166135741 \\
84 & $3d5d$ & -2.9024 0839 2145 & 2.00023 & -0.000000020082 \\
85 & $4d5d$ & -2.9024 8149 3317 & 2.00015 & -0.000073101172 \\
86 & $5d5d$ & -2.9026 2291 3640 & 2.00010 & -0.000141420323 \\
87 & $3d6d$ & -2.9026 2603 4370 & 2.00009 & -0.000003120730 \\                       
88 & $4d6d$ & -2.9026 2788 7514 & 2.00010 & -0.000001853144 \\
89 & $5d6d$ & -2.9026 2848 6653 & 2.00009 & -0.000000599139 \\                      
90 & $6d6d$ & -2.9026 5476 6624 & 2.00007 & -0.000026279971 \\                    
91 & $3d7d$ & -2.9026 5479 4554 & 2.00007 & -0.000000027930 \\                     
92 & $4d7d$ & -2.9026 5499 2659 & 2.00007 & -0.000000198106 \\                    
93 & $5d7d$ & -2.9026 6153 4641 & 2.00006 & -0.000006541983 \\                       
94 & $6d7d$ & -2.9026 6505 9330 & 2.00006 & -0.000003524688 \\                      
95 & $7d7d$ & -2.9026 7927 5292 & 2.00005 & -0.000014215962 \\                     
96 & $3d8d$ & -2.9026 7947 5963 & 2.00005 & -0.000000200671 \\                         
97 & $4d8d$ & -2.9026 7986 2235 & 2.00005 & -0.000000386272 \\                       
98 & $5d8d$ & -2.9026 8195 5846 & 2.00005 & -0.000002093611 \\
99 & $6d8d$ & -2.9026 9330 7132 & 2.00003 & -0.000011351286 \\                          
100 & $7d8d$ & -2.9026 9351 9452 & 2.00003 & -0.000000212320 \\
101 & $8d8d$ & -2.9027 0018 1175 & 2.00002 & -0.000006661723 \\
\hline
Exact \cite{Kurokawa} &  & -2.9037 2437 7034 1195 9831 &   & -0.001024195859   \\ 
\hline\hline
\end{tabular}
\end{center}
\end{table}

\clearpage



\begin{turnpage}

\begin{landscape}
\begin{table}
\caption{Comparison of extensive $s$, $p$, $d$-CI calculations on the $1^1S$ ground 
of helium atom. } 
\begin{tabular}{ccccccrcrc}
\hline\hline
& $L_{\rm max}$ & $N_{\rm max}$ & Confs. & This work$^a$ &$N_{\rm max}$ & Confs. &   Sims and Hagstrom$^b$ \cite{Sims02} 
&$N_{\rm orb}$ & Bromley and Mitroy$^c$ \cite{Bromley}  \\
&  0 ($s$-limit) & 15 & 296 & -2.8790 2868 09 & 21 & 470  & -2.8790 2875 65 & 44 &-2.8790 2876 0   \\
&  1 ($p$-limit) & 16 & 497 & -2.9005 1585 44 & 21 & 854  & -2.9005 1621 99 & 80 &-2.9005 1622 8  \\
&  2 ($d$-limit) & 16 & 668 & -2.9027 6626 64 & 21 & 1221 & -2.9027 6680 53 &115 &-2.9027 6682 3  \\
{\rm Best CI} & 18 ($v$-limit) & 15 &    &  &      & 4699 & -2.9037 200 919 & 465 & -2.9037 1278 6 \\
\hline\hline
\end{tabular}
\footnotetext[1]{In this work the exponents of Table II were used.} 
\footnotetext[2]{Sims and Hagstrom used $4.10$ and $25.0$ for $s$-orbitals, $3.05$ and $40.5$ for 
$p$-orbitals and $3.50$ and $40.5$ for $d$-orbitals. }
\footnotetext[3]{Bromley and Mitroy used Laguerre functions with the exponent $8.6$ for $s$-orbitals, 
11.6 for $p$-orbitals and 14.4 for $d$-orbitals. $N_{\rm orb}$ is the number of orbitals.}  
\end{table}
\end{landscape}

\end{turnpage}

\clearpage

\newpage


\begin{table}[tp]
\begin{center}
\caption{s, p and d-Hy-CI calculations on the $1^1S$ ground state of helium 
atom using the basis of optimized orbital exponents of Table II. The calculation time was 87 seconds in our 
computer using QP. The achieved accuracy is of $\approx$ 0.5 microhartree.}
\begin{tabular}{ccccc}
\hline\hline
N & Configuration \quad & \quad Energy ($E_h$) \qquad & \qquad Virial \qquad & \qquad 
Difference ($E_h$) \\ 
\hline
1 & $1s1s$ & -2.8046 0000 0000 0000 0000 & 2.280405 & 0.0$\;\;\;\;\;\;\;\;\;\;\;\;\;\;\;\;\;\;\;$ \\ 
2 & $1s2s$ & -2.8464 5649 2448 7886 8848 & 2.012655 & -0.041856492449 \\ 
3 & $2s2s$ & -2.8528 0522 2662 7660 6893 & 2.038758 & -0.006348730214 \\ 
4 & $1s1s^{\prime}$ & -2.8749 7028 4518 6921 8631 & 2.002043 & -0.022165061856 \\ 
5 & $2s1s^{\prime}$ & -2.8753 1452 7730 4383 9933 & 2.001306 & -0.000344243212 \\ 
6 & $1s^{\prime}1s^{\prime}$ & -2.8762 1971 7810 5121 5735 & 2.000400 & -0.000905190080 \\ 
7 & $1s2s^{\prime}$ & -2.8774 6546 4594 0870 3077 & 2.001049 & -0.001245746784 \\ 
8 & $2s2s^{\prime}$ & -2.8784 4308 9302 1401 8945 & 2.000063 & -0.000977624708 \\ 
9 & $1s^{\prime}2s^{\prime}$ & -2.8784 9065 1588 5673 6622 & 2.000045 & -0.000047562286 \\ 
10 & $2s^{\prime}2s^{\prime}$ & -2.8787 4806 7775 6347 2228 & 1.999859 & -0.000257416187 \\ 
11 & $1s3s$ & -2.8787 8660 6389 3286 6871 & 2.000036 & -0.000038538614 \\ 
12 & $2s3s$ & -2.8788 2609 4249 0916 2944 & 2.000087 & -0.000039487860 \\ 
13 & $1s^{\prime}3s$ & -2.8789 0234 0012 1737 8000 & 2.000025 & -0.000076245763 \\ 
14 & $2s^{\prime}3s$ & -2.8789 0276 7327 8845 3632 & 2.000022 & -0.000000427316 \\ 
15 & $3s3s$ & -2.8789 5525 8603 6072 3426 & 1.999999 & -0.000052491276 \\ 
16 & $1s1sr_{12}$ & -2.9030 3855 5047 6765 9888 & 2.000060 & -0.024083296444 \\ 
17 & $1s2sr_{12}$ & -2.9030 3963 0490 7037 1147 & 1.999905 & -0.000001075443 \\ 
18 & $2s2sr_{12}$ & -2.9032 5861 2556 4860 4038 & 2.000030 & -0.000218982066 \\ 
19 & $1s1s^{\prime}r_{12}$ & -2.9033 8299 6275 5132 6846 & 2.000091 & -0.000124383719 \\ 
20 & $2s1s^{\prime}r_{12}$ & -2.9034 1178 0236 9870 1758 & 2.000012 & -0.000028783961 \\ 
21 & $1s^{\prime}1s^{\prime}r_{12}$ & -2.9034 3722 4940 8831 4415 & 1.999989 & -0.000025444704 \\ 
22 & $1s2s^{\prime}r_{12}$ & -2.9034 3950 8142 1374 1002 & 1.999972 & -0.000002283201 \\ 
23 & $2s2s^{\prime}r_{12}$ & -2.9034 6815 0498 5144 6232 & 1.999956 & -0.000028642356 \\ 
24 & $1s^{\prime}2s^{\prime}r_{12}$ & -2.9034 6820 5099 9614 1845 & 1.999956 & -0.000000054601 \\ 
25 & $2s^{\prime}2s^{\prime}r_{12}$ & -2.9034 7311 4551 4696 4381 & 1.999952 & -0.000004909451 \\ 
26 & $1s3sr_{12}$ & -2.9034 7648 8273 0412 1726 & 1.999992 & -0.000003373722 \\ 
27 & $2s3sr_{12}$ & -2.9034 7957 2554 5670 0172 & 2.000002 & -0.000003084282 \\ 
\hline\hline 
\end{tabular}
\end{center}
\end{table}

\newpage


\begin{table}[tp]
\begin{center}
\textbf{Table V} Continuation. 
\begin{tabular}{ccccc}
\hline\hline
N & Configuration \quad & \quad Energy ($E_h$) \qquad & \qquad Virial \qquad & \qquad Difference ($E_h$) \\
\hline
28 & $1s^{\prime}3sr_{12}$ & -2.9034 8135 6723 1881 1971 & 2.000003 & -0.000001784169 \\
29 & $2s^{\prime}3sr_{12}$ & -2.9034 8172 3768 1817 9421 & 2.000002 & -0.000000367045 \\
30 & $3s3sr_{12}$ & -2.9034 8191 0228 3344 9701 & 2.000003 & -0.000000186460 \\
31 & $2p2p$ & -2.9036 9490 9181 3235 6472 & 1.999946 & -0.000212998953 \\
32 & $2p3p$ & -2.9037 0333 9014 5345 8498 & 1.999973 & -0.000008429833 \\
33 & $3p3p$ & -2.9037 1382 9571 8228 7221 & 1.999986 & -0.000010490557 \\
34 & $2p2p^{\prime}$ & -2.9037 1407 2375 9450 1950 & 1.999985 & -0.000000242804 \\
35 & $3p2p^{\prime}$ & -2.9037 1429 5379 2885 5478 & 1.999989 & -0.000000223003 \\
36 & $2p^{\prime}2p^{\prime}$ & -2.9037 1572 6120 9390 5519 & 1.999990 & -0.000001430742 \\
37 & $2p3p^{\prime}$ & -2.9037 1588 5177 7166 9859 & 1.999993 & -0.000000159057 \\
38 & $3p3p^{\prime}$ & -2.9037 1634 2690 0297 1140 & 1.999994 & -0.000000457512 \\
39 & $2p^{\prime}3p^{\prime}$ & -2.9037 1644 9779 9809 6406 & 1.999993 & -0.000000107090 \\
40 & $3p^{\prime}3p^{\prime}$ & -2.9037 1644 9785 9439 7737 & 1.999993 & -0.000000000006 \\ 
41 & $2p2pr_{12}$ & -2.9037 2173 2417 9137 9221 & 1.999989 & -0.000005282632 \\ 
42 & $2p3pr_{12}$ & -2.9037 2182 0699 2263 4691 & 1.999990 & -0.000000088281 \\ 
43 & $3p3pr_{12}$ & -2.9037 2214 9854 2463 6331 & 1.999993 & -0.000000329155 \\ 
44 & $2p2p^{\prime}r_{12}$ & -2.9037 2227 8371 0330 7986 & 1.999994 & -0.000000128517 \\ 
45 & $3p2p^{\prime}r_{12}$ & -2.9037 2237 5963 4585 9805 & 1.999994 &  -0.000000097592 \\ 
46 & $2p^{\prime}2p^{\prime}r_{12}$ & -2.9037 2265 5745 0023 6384 & 1.999994 & -0.000000279782 \\ 
47 & $2p3p^{\prime}r_{12}$ & -2.9037 2272 5043 9425 8433 & 1.999994 & -0.000000069299 \\ 
48 & $3p3p^{\prime}r_{12}$ & -2.9037 2278 0206 3073 0904 & 1.999995 & -0.000000055162 \\ 
49 & $2p^{\prime}3p^{\prime}r_{12}$ & -2.9037 2287 1724 9868 5604 & 1.999996 & -0.000000091519 \\ 
50 & $3p^{\prime}3p^{\prime}r_{12}$ & -2.9037 2311 0207 5489 3605 & 1.999997 & -0.000000238483 \\ 
51 & $3d3d$ & -2.9037 2311 5896 1644 4399 & 1.999997 & -0.000000005689 \\ 
52 & $3d4d$ & -2.9037 2325 3694 3809 8700 & 1.999996 & -0.000000137798 \\ 
53 & $4d4d$ & -2.9037 2341 0977 7800 8564 & 1.999997 & -0.000000157283 \\ 
54 & $3d5d$ & -2.9037 2363 2223 7418 5292 & 1.999998 & -0.000000221246 \\ 
55 & $4d5d$ & -2.9037 2368 3956 6486 0703 & 1.999998 & -0.000000051733 \\ 
56 & $5d5d$ & -2.9037 2371 6568 3659 1072 & 1.999999 & -0.000000032612 \\ 
\hline\hline
\end{tabular}
\end{center}
\par
\end{table}

\newpage


\begin{table}[tp]
\begin{center}
\textbf{Table V} Continuation.
\begin{tabular}{ccccc}
\hline\hline
N & Configuration \quad & \quad Energy ($E_h$) \qquad & \qquad Virial \qquad & 
\qquad Difference ($E_h$)\\
\hline
7 & $3d3dr_{12}$ & -2.9037 2375 1435 0361 1170 & 1.999999 & -0.000000034867 \\ 
58 & $3d4dr_{12}$ & -2.9037 2375 1898 7744 7923 & 1.999999 & -0.000000000464 \\ 
59 & $4d4dr_{12}$ & -2.9037 2376 3066 1777 0715 & 1.999999 & -0.000000011167 \\ 
60 & $3d5dr_{12}$ & -2.9037 2377 6768 3902 6301 & 1.999999 & -0.000000013702 \\ 
61 & $4d5dr_{12}$ & -2.9037 2377 8464 9764 7830 & 1.999999 & -0.000000001697 \\ 
62 & $5d5dr_{12}$ & -2.9037 2378 4817 0105 8051 & 1.999999 & -0.000000006352 \\ 
\hline
Exact \cite{Kurokawa} &  & -2.9037 2437 7034 1195 9831 &  & -0.000000592217 \\ 
\hline\hline
\end{tabular}
\end{center}
\end{table}

\newpage



\begin{table}[tp]
\begin{center}
\caption{Hy-CI calculation on the $1^1S$ state of the helium atom using s-, p-, d-, and 
f-orbitals and the single orbital exponent per shell $\alpha = 2.98140$ for all configurations. 
The notation '1:9s 1:9s' stays for a block of configurations
constructed systematically with the orbital set $1s,2s,3s,4s,5s,6s,7s,8s,9s$. 
The calculation time was 4408 seconds in our computer using QP. The achieved accuracy is of $\approx$ 22 picohartree.}
\begin{tabular}{ccrrcr}
\hline\hline
Conf. & Wave function \quad & \quad N \quad & \quad N$_{tot}$ \quad & \quad E ($E_h$) \quad & \quad Difference in $\mu E_h$ \\
\hline
ss         & 1:9s 1:9s    & 45  & 45  & -2.8790 1861 6029 5731   & 0.0$\;\;\;\;\;\;\;\;\;\;\;\;\;\;\;\;$ \\
ss$r_{12}$ & 1:9s 1:9s    & 45  & 90  & -2.9034 9650 6178 4873 & -24477.8901489142 \\
pp         & 2:9p 2:9p    & 36  & 126 & -2.9037 1950 0968 3934   & -222.9947899061  \\
pp$r_{12}$ & 2:9p 2:9p    & 36  & 162 & -2.9037 2403 4569 5773   &   -4.5336011839  \\ 
dd         & 3:9d 3:9d    & 28  & 190 & -2.9037 2415 3444 2346   &   -0.1188746573  \\
dd$r_{12}$ & 3:9d 3:9d    & 28  & 218 & -2.9037 2415 8630 9900   &   -0.0051867554  \\
ff         & 4:8f 4:8f    & 21  & 239 & -2.9037 2415 9128 6842   &   -0.0004976942  \\
ff$r_{12}$ & 4:8f 4:8f    & 21  & 260 & -2.9037 2415 9246 2237   &   -0.0001175395  \\
ss         & 1:14s 10:14s & 60  & 320 & -2.9037 2437 6412 0673   &   -0.2171658436  \\
ss$r_{12}$ & 1:14s 10:14s & 60  & 380 & -2.9037 2437 6847 4482   &   -0.0004353809  \\    
pp         & 2:14p 10:14p & 55  & 435 & -2.9037 2437 6891 4037   &   -0.0000439555  \\
pp$r_{12}$ & 2:14p 10:14p & 55  & 490 & -2.9037 2437 6920 6840   &   -0.0000292803  \\ 
dd         & 3:14d 10:14d & 50  & 540 & -2.9037 2437 6927 2929   &   -0.0000066089  \\
dd$r_{12}$ & 3:14d 10:14d & 50  & 590 & -2.9037 2437 6929 6796   &   -0.0000023867  \\
ff         & 4:14f 10:14f & 45  & 635 & -2.9037 2437 6930 0411   &   -0.0000003615  \\
ff$r_{12}$ & 4:14f 10:14f & 45  & 680 & -2.9037 2437 6930 1280   &   -0.0000000869  \\
ss         & 1:18s 15:18s & 66  & 746 & -2.9037 2437 6994 8804   &   -0.0000647524  \\
ss$r_{12}$ & 1:18s 15:18s & 66  & 812 & -2.9037 2437 7010 1954   &   -0.0000153150  \\
pp         & 2:16p 15:16p & 29  & 841 & -2.9037 2437 7010 3515   &   -0.0000001561  \\
pp$r_{12}$ & 2:16p 15:16p & 29  & 870 & -2.9037 2437 7011 2489   &   -0.0000008974  \\
dd         & 3:16d 15:16d & 27  & 897 & -2.9037 2437 7011 2646   &   -0.0000000157  \\
dd$r_{12}$ & 3:16d 15:16d & 27  & 924 & -2.9037 2437 7011 8739   &   -0.0000006093  \\
ff         & 4:16f 15:16d & 25  & 949 & -2.9037 2437 7011 9468   &   -0.0000000729  \\
ff$r_{12}$ & 4:16f 15:16d & 25  & 974 & -2.9037 2437 7011 9675   &   -0.0000000207  \\  
\hline 
Exact \cite{Kurokawa} &  &   &       & -2.9037 2437 7034 1195 9831   &   -0.0000221520 \\ 
\hline\hline
\end{tabular}
\end{center}
\end{table}
%



\newpage


\begin{table}[tp]
\begin{center}
\caption{Hy calculations on the $1^1S$ ground state of the helium atom using s-orbitals 
and $r_{12}^{\nu}$ with $\nu \le 9$ and a single orbital exponent per shell $\alpha = 2.98140$ for all configurations. 
N and N$_{tot}$ are the number of configurations employed. Calculation time was 8275 seconds in our computer using QP. 
An accuracy of $\approx$ 3 picohartree has been achieved.}
\scalebox{0.80}{
\begin{tabular}{ccrrcr}
\hline\hline
Configuration & Wave function \quad & \quad N \quad & \quad N$_{tot}$ \quad & \quad Energy ($E_h$)  & Difference in $\mu E_h$ \\
\hline
ss           & 1:9s 1:9s    & 45  & 45  & -2.8790 1861 6029 5731  &  0.0$\;\;\;\;\;\;\;\;\;\;\;\;\;\;\;\;$ \\
ss$r_{12}$   & 1:9s 1:9s    & 45  & 90  & -2.9034 9650 6178 4873  & -24477.8901489142        \\
ss$r_{12}^2$ & 1:9s 1:9s    & 45  & 135 & -2.9037 1969 0132 0298    & -223.1839535425        \\
ss$r_{12}^3$ & 1:9s 1:9s    & 45  & 180 & -2.9037 2423 2335 0404      & -4.5422030106         \\
ss$r_{12}^4$ & 1:9s 1:9s    & 45  & 225 & -2.9037 2436 3833 1531      & -0.1314981127         \\
ss$r_{12}^5$ & 1:9s 1:9s    & 45  & 270 & -2.9037 2437 1731 5418      & -0.0078983887         \\
ss$r_{12}^6$ & 1:9s 1:9s    & 45  & 315 & -2.9037 2437 3803 9327      & -0.0020723909         \\
ss$r_{12}^7$ & 1:9s 1:9s    & 45  & 360 & -2.9037 2437 4635 1568      & -0.0008312241         \\
ss$r_{12}^8$ & 1:9s 1:9s    & 45  & 405 & -2.9037 2437 5061 4324      & -0.0004262756         \\
ss$r_{12}^9$ & 1:9s 1:9s    & 45  & 450 & -2.9037 2437 5315 1465      & -0.0002537141         \\
ss           & 1:14s 10:14s & 60  & 510 & -2.9037 2437 6988 5853      & -0.0016734388         \\
ss$r_{12}$   & 1:14s 10:14s & 60  & 570 & -2.9037 2437 6993 4106      & -0.0000048253         \\
ss$r_{12}^2$ & 1:14s 10:14s & 60  & 630 & -2.9037 2437 6996 1059      & -0.0000026953         \\
ss$r_{12}^3$ & 1:14s 10:14s & 60  & 690 & -2.9037 2437 6998 2144      & -0.0000021085         \\
ss$r_{12}^4$ & 1:14s 10:14s & 60  & 750 & -2.9037 2437 7000 8505      & -0.0000026361         \\
ss$r_{12}^5$ & 1:14s 10:14s & 60  & 810 & -2.9037 2437 7011 4174      & -0.0000105669         \\
ss$r_{12}^6$ & 1:14s 10:14s & 60  & 870 & -2.9037 2437 7017 9404      & -0.0000065230         \\
ss$r_{12}^7$ & 1:14s 10:14s & 60  & 930 & -2.9037 2437 7022 8414      & -0.0000049010         \\
ss$r_{12}^8$ & 1:14s 10:14s & 60  & 990 & -2.9037 2437 7026 0986      & -0.0000032572         \\
ss$r_{12}^9$ & 1:14s 10:14s & 60  &1050 & -2.9037 2437 7028 3340      & -0.0000022354         \\
ss           & 1:16s 15:19s & 85  &1135 & -2.9037 2437 7028 3466      & -0.0000000126         \\
ss$r_{12}$   & 1:16s 15:16s & 31  &1166 & -2.9037 2437 7028 3474      & -0.0000000008         \\
ss$r_{12}^2$ & 1:16s 15:16s & 31  &1197 & -2.9037 2437 7028 3483      & -0.0000000009         \\
ss$r_{12}^3$ & 1:16s 15:16s & 31  &1228 & -2.9037 2437 7028 3525      & -0.0000000042         \\
ss$r_{12}^4$ & 1:16s 15:16s & 31  &1259 & -2.9037 2437 7028 3545      & -0.0000000020         \\
ss$r_{12}^5$ & 1:16s 15:16s & 31  &1290 & -2.9037 2437 7028 3960      & -0.0000000415         \\
ss$r_{12}^6$ & 1:16s 15:16s & 31  &1321 & -2.9037 2437 7028 5055      & -0.0000001095         \\
ss$r_{12}^7$ & 1:16s 15:16s & 31  &1352 & -2.9037 2437 7028 6388      & -0.0000001333         \\
ss$r_{12}^8$ & 1:16s 15:16s & 31  &1383 & -2.9037 2437 7029 9987      & -0.0000013599         \\
ss$r_{12}^9$ & 1:16s 15:16s & 31  &1414 & -2.9037 2437 7031 0478      & -0.0000010491         \\
\hline
Exact \cite{Kurokawa} & & &            & -2.9037 2437 7034 1195 9831      & -0.0000030718    \\
\hline\hline
\end{tabular}
}
\end{center}
\end{table}

\newpage



\begin{table}[tp]
\begin{center}
\caption{Pattern of convergence of the Hy calculation on the $1^1S$ ground state of the 
helium atom using s-orbitals and $r_{12}^{\nu}$ with $\nu \le 8$ and a single orbital exponent $\alpha = 2.918780$ 
per shell for all configurations. The blocks of configurations are ordered by the power $\nu$. The calculation time was 520 seconds 
in our computer using QP. An accuracy of $\approx$ 20 picohartree has been achieved. }
\begin{tabular}{lcrrcr}
\hline\hline
Conf. & Wave function & N & N$_{tot}$ & Energy ($E_h$) & Difference in $\mu E_h$ \\
\hline
ss           & 1:19s 1:19s  & 190 & 190 & -2.8790 2777 3182 8171  & 0.0$\;\;\;\;\;\;\;\;\;\;\;\;\;\;\;\;$ \\
ss$r_{12}$   & 1:12s 1:12s  &  78 & 268 & -2.9034 9776 5241 2314  & -24469.9920584143    \\
ss$r_{12}^2$ & 1:12s 1:12s  &  78 & 346 & -2.9037 1985 0441 3351  &   -222.0852001037     \\
ss$r_{12}^3$ & 1:12s 1:12s  &  78 & 424 & -2.9037 2426 5424 1621  &     -4.4149828270     \\
ss$r_{12}^4$ & 1:12s 1:12s  &  78 & 502 & -2.9037 2437 3962 4616  &     -0.1085382995     \\
ss$r_{12}^5$ & 1:12s 1:12s  &  78 & 580 & -2.9037 2437 6911 8283  &     -0.0029493667     \\
ss$r_{12}^6$ & 1:12s 1:12s  &  78 & 658 & -2.9037 2437 7002 4570  &     -0.0000906287     \\
ss$r_{12}^7$ & 1:12s 1:12s  &  78 & 736 & -2.9037 2437 7008 0749  &     -0.0000056179     \\
ss$r_{12}^8$ & 1:12s 1:12s  &  78 & 814 & -2.9037 2437 7015 4499  &     -0.0000073750     \\
\hline
Exact \cite{Kurokawa}  &   &   &   & -2.9037 2437 7034 1195 9831 &     -0.0000186697 \\
\hline\hline
\end{tabular}
\end{center}
\end{table}

\newpage



\begin{table}[tp]                  
\begin{center}
\caption{Comparison of extensive CI-R12, Hy, Hy-CI calculations on the $1^1$S state of 
helium atom which were done using the same basis set, same computer program and same 
computer machine. In this work the same single orbital exponent 2.9814000 
was used in the Hy, Hy-CI and CI-R12 calculations. These values are obtained in QP numerical precision.}
\begin{tabular}{ccccrcrr}
\hline\hline
Method & $L_{\rm max}$ & $N_{\rm max}$ & $\nu$ & Confs. &  Energy ($E_h$)  & Time(s) & Accuracy \\
\hline
 Hy-CI/CI-R12$^a$ & 0 & 20 & 1 & 420   & -2.9034 9813 3768 2647 9878  & 138  & 0.2 {\rm  milihartree}\\    
 Hy-CI & 1 & 20 & 1 & 800   & -2.9037 2426 7864 5688 1088  & 1356 & 0.1 {\rm microhartree} \\             
 Hy-CI & 2 & 20 & 1 & 1142  & -2.9037 2437 6943 7414 6155  & 4811 & 0.1 {\rm nanohartree} \\ 
 Hy-CI & 3 & 16 & 1 & 939   & -2.9037 2437 7011 9453 9894  & 4302 & 22  {\rm picohartree} \\ 
 Hy    & 0 & 12 & 8 & 814   & -2.9037 2437 7018 2197 9713  & 1160 & 16 {\rm picohartree} \\
 Hy-CI & 3 & 20 & 1 & 1452  & -2.9037 2437 7024 3236 8211  & 6600 & 10 {\rm picohartree} \\           
 Hy$^b$    & 0 & 16 & 9 & 1414  & -2.9037 2437 7031 0477 6022  & 8788 &  3 {\rm picohartree} \\ 
\hline
{\rm Exact}\cite{Kurokawa} &  & & &  & -2.9037 2437 7034 1195 9831 12 &      &   \\  
\hline\hline
\end{tabular}
\end{center}
\footnotetext[1]{Note that the CI-R12 method including excitations others than the ones 
of the orbitals occurring in the ground state configuration is equivalent for the two-electron case to the Hy-CI wave function.}
\footnotetext[2]{Hy configurations with $L \ge 16$ suffer from linear dependence problems in our computer. Higher numerical precision than QP is needed.} 
\end{table}

\end{document}